\documentclass[a4paper,11pt]{article}
\usepackage{graphicx,caption}
\usepackage{gensymb}
\usepackage{subfigure}
\usepackage{bigints}
\graphicspath{{Figures/}}
\usepackage{lineno}
\usepackage{subfigure}

\pdfoutput=1 

\usepackage{jinstpub} 

\title{\boldmath Experimental Determination of Proton Hardness Factors at Several Irradiation Facilities}

\author[a]{P.~Allport,}
\author[c]{F.~B\"ogelspacher,}
\author[a]{K.~Bruce,}
\author[a]{R.~Canavan,}
\author[c]{A.~Dierlamm,}
\author[a]{L.~Gonella,}
\author[a]{P.~Knights,}
\author[b]{I.~Mateu,}
\author[b]{M.~Moll,}
\author[a]{K.~Nikolopoulos,}
\author[a]{B.~Phoenix,}
\author[a]{T.~Price,}
\author[a]{L.~Ram,}
\author[b]{F.~Ravotti,}
\author[a]{C.~Simpson-Allsop,}
\author[a]{and C.~Wood}

\affiliation[a]{University of Birmingham,\\ Birmingham, B15 2TT, United Kingdom}
\affiliation[b]{IRRAD Proton Facility,\\CERN, CH-1211 Geneva 23, Switzerland}
\affiliation[c]{Institute of Experimental Particle Physics, Karlsruhe Institute of Technology,\\ Karlsruhe, D-76131, Germany}

\emailAdd{k.nikolopoulos@bham.ac.uk}

\abstract{The scheduled High Luminosity upgrade of the CERN Large Hadron
Collider presents new challenges in terms of radiation hardness. As a
consequence, campaigns to qualify the radiation hardness of detector
sensors and components are undertaken worldwide. The effects of
irradiation with beams of different particle species and energy,
aiming to assess displacement damage in semiconductor devices, are
communicated in terms of the equivalent $1\;$MeV neutron fluence,
using the hardness factor for the conversion.  In this work, the
hardness factors for protons at three different kinetic energies have
been measured by analysing the I--V and C--V characteristics of
reverse biased diodes, pre- and post-irradiation.  The sensors were
irradiated at the MC40 Cyclotron of the University of Birmingham, the
cyclotron at the Karlsruhe Institute of Technology, and the IRRAD
proton facility at CERN, with the respective measured proton hardness
factors being: 
$2.1\pm 0.5$ for $24\;$MeV, 
$2.2 \pm 0.4$ for $23\;$MeV, and $0.62\pm 0.04$ for $23\;$GeV. The hardness factors
currently used in these three facilities are in agreement with the
presented measurements.
}

\begin{document}
\maketitle
\flushbottom
\newpage
\section{Introduction}
The scheduled upgrade of the CERN Large Hadron
Collider~\cite{Evans:2008zzb} to its High Luminosity phase (HL--LHC)
in 2024~\cite{Bruning:2015dfu} presents new challenges in terms of
detector radiation
hardness~\cite{ATLASLoI,CMSCollaboration:2015zni}. In irradiation
facilities around the world, campaigns to qualify radiation hardness
of detector sensors are undertaken. Standard $1\;$MeV neutron
equivalent fluences are used for the purpose of inter-facility
comparison and collaboration. The experimentally determined proton
fluences at given energies are converted to the standardised fluences
using the corresponding hardness factor. This factor is usually
derived by evaluating the leakage current in the bulk of a silicon
sensor; assuming that the measured leakage current is scaling with the
non-ionizing energy loss (NIEL).

At the University of Birmingham, the MC40 cyclotron adopts a hardness
factor value of 2.2~\cite{Dervan:2015pua} for protons of
$23\;$MeV. However, for beams of similar kinetic energy, facilities
have obtained different values. For example, the Irradiation Center
Karlsruhe has recorded a value of $2.05 \pm 0.61$ for $24\;$MeV
protons, with a previously assumed value of 1.85 for $26\;$MeV
protons~\cite{Karlsruhe}. Other studies have claimed similar values,
such as 2.22 for $27\;$MeV protons~\cite{Moll:2002tn}. Tabulated
values from the RD50 Collaboration, which is concerned with the
development of radiation hard semiconductor devices for high
luminosity colliders~\cite{RD50}, give a value of approximately $2.56$
for $25\;$MeV protons~\cite{Huhtinen:1993np}. For $23\;$GeV protons, a
value of 0.62 has been measured at the IRRAD proton
facility~\cite{Moll:2002tn}.  Due to these discrepancies, further
study on the value of the hardness factor is required. Moreover, the
large uncertainties of these measurements result to large
uncertainties when the proton fluences are converted to the
corresponding $1\;$MeV neutron equivalent fluences.

In this article the current-voltage (I--V) and capacitance-voltage
(C--V) characteristics of reversed biased BPW34F photodiodes are
analysed to obtain proton hardness factors for protons of three
different kinetic energies. In the following the irradiation
facilities are presented briefly in Sec.~\ref{sec:irradiations}, while
the description of the measurements is given in
Sec.~\ref{sec:measurements}. The obtained results are presented in
Sec.~\ref{sec:results}, while a concluding discussion is given in
Sec.~\ref{sec:summary}.

\section{Irradiations}
\label{sec:irradiations}
\subsection{MC40 Cyclotron}
\begin{figure}[htbp]
  \centering 
  \subfigure[\label{fig:ATLASChamberLeft}]{\includegraphics[width=.35\textwidth]{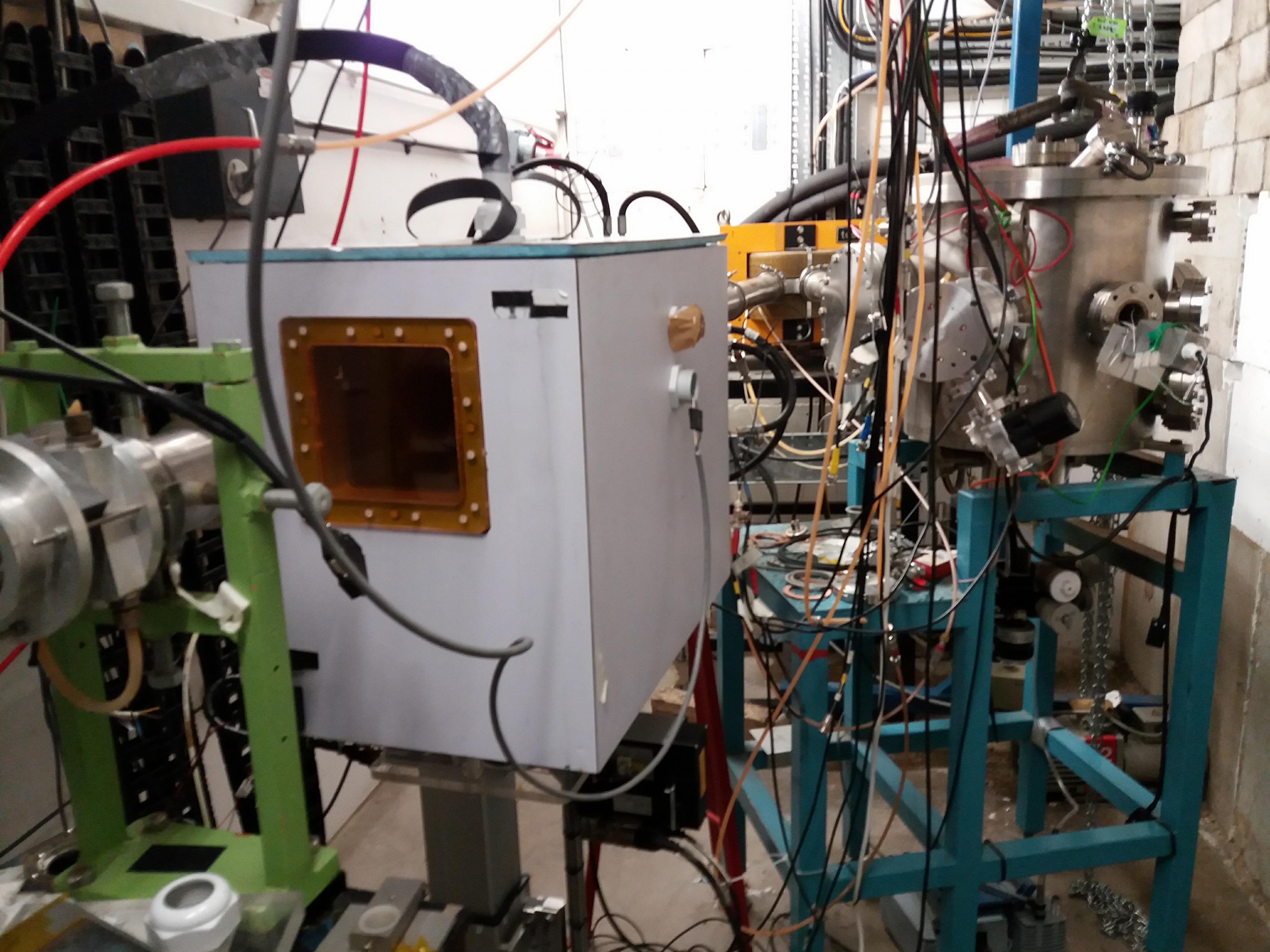}}
  \subfigure[\label{fig:ATLASChamberRight}]{\includegraphics[width=.35\textwidth]{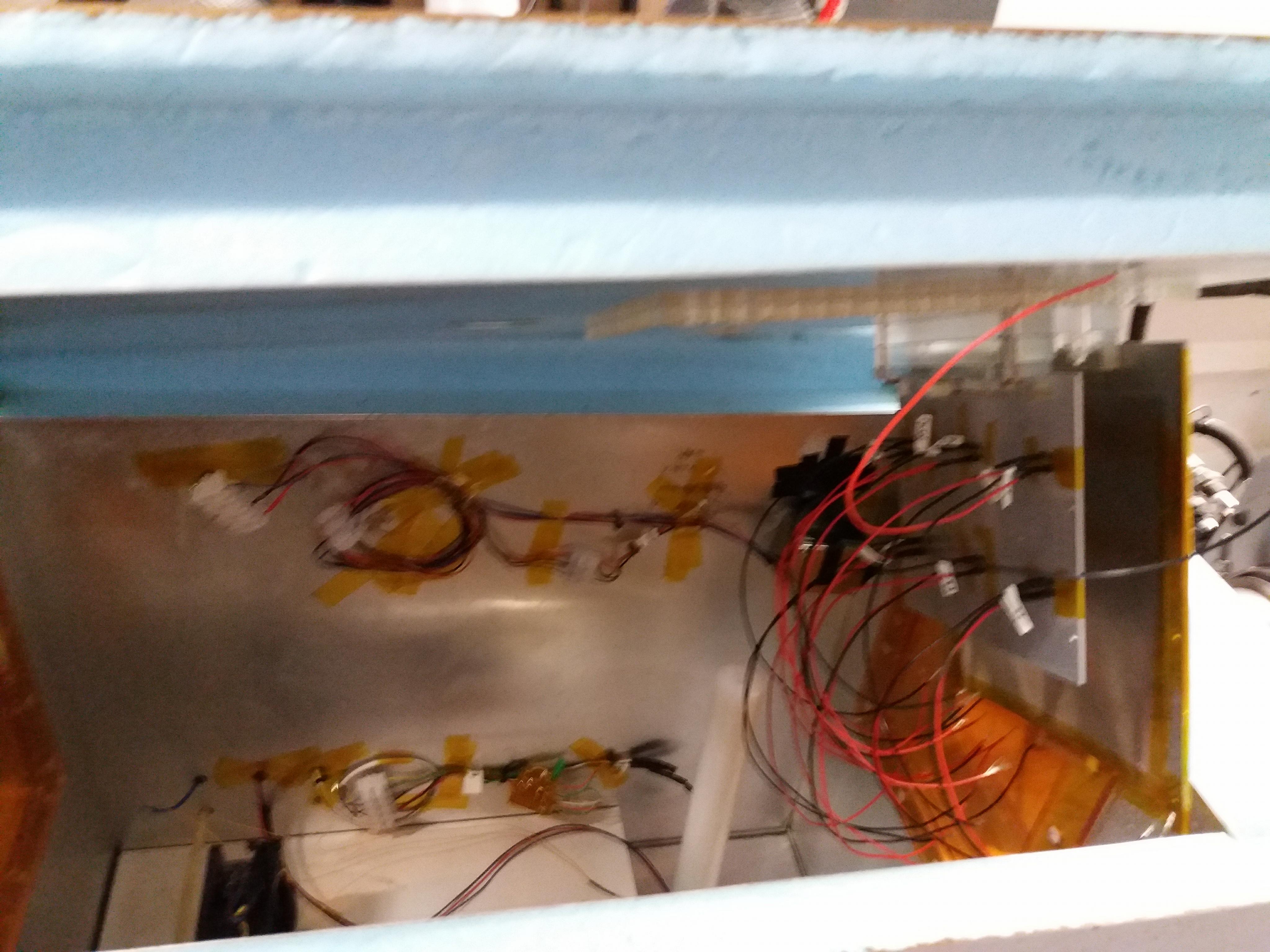}}
  \caption{\subref{fig:ATLASChamberLeft} The high intensity area of the MC40 cyclotron with the temperature controlled chamber. \subref{fig:ATLASChamberRight} The interior of the temperature controlled chamber, viewed from the side. The aluminium plate used to mount the diodes is visible on the right.\label{fig:ATLASChamber}}
\end{figure}

The MC40 cyclotron at the University of Birmingham is primarily used
for the production of medical isotopes. However, it is regularly used
for nuclear physics research and radiation damage studies. The
configuration utilised for high intensity proton irradiations is shown
in Fig.~\ref{fig:ATLASChamber}. 
The beam spot is a square of area $10\times10\;$mm\textsuperscript{2},
and its position is calibrated before each irradiation session with
gafchromic film. The sample is isolated from the environment using a
temperature controlled chamber, which is is mounted on a XY-axis
robotic scanning system controlled via LabView. The chamber
temperature during irradiation is set to $-27\degree$C, this is
selected to ensure that the sample temperature remains well below $0\degree$C,
even when irradiated at the highest available dose rate in the
facility, and thus there will be no significant contribution to
thermal annealing.
Further details for the
irradiation facility are provided in Ref.~\cite{Allport:2017ipp}.

\begin{figure}[htbp]
  \centering 
  \subfigure[\label{fig:DiodeMountLeft}]{\includegraphics[width=.35\textwidth]{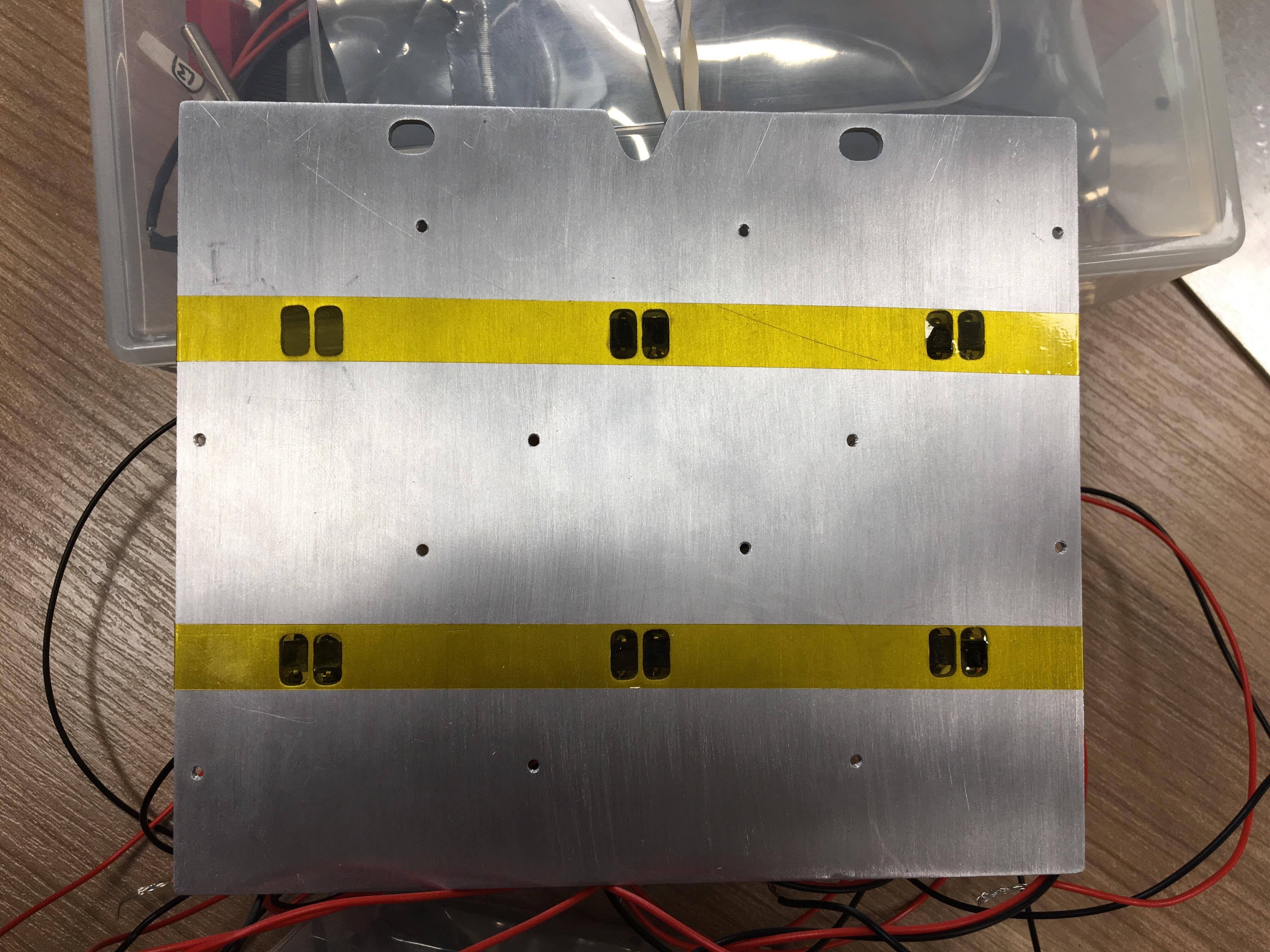}}
  \subfigure[\label{fig:DiodeMountRight}]{\includegraphics[width=.35\textwidth]{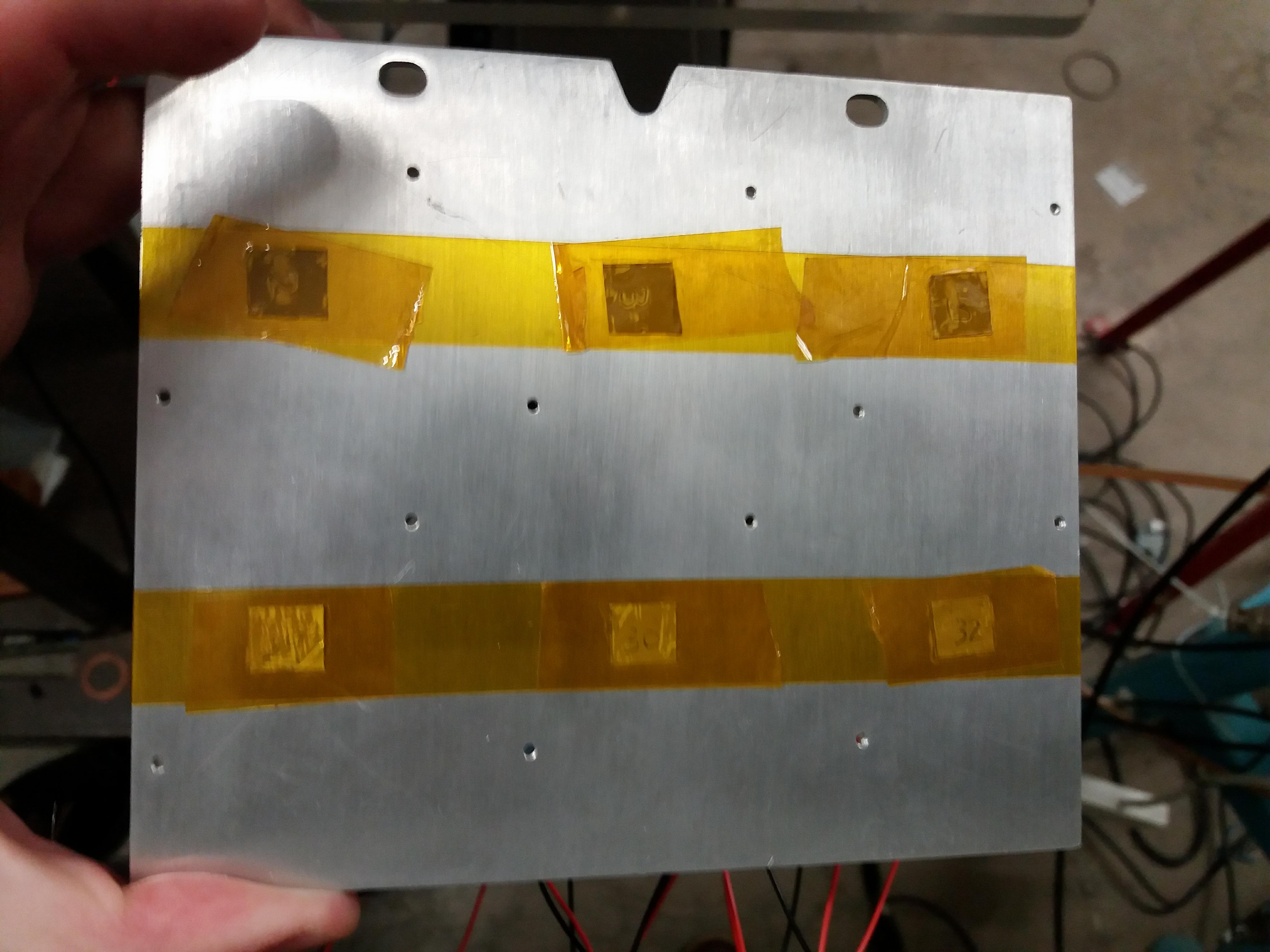}}
  \caption{\subref{fig:DiodeMountLeft} Aluminium diode mount with attached diodes; and \subref{fig:DiodeMountRight} the same mount following placement of ${}^{57}$Ni foils for fluence measurements.\label{fig:DiodeMount}}
    \end{figure}

\begin{figure}[htbp]
  \centering
  \includegraphics[width = 0.75\linewidth]{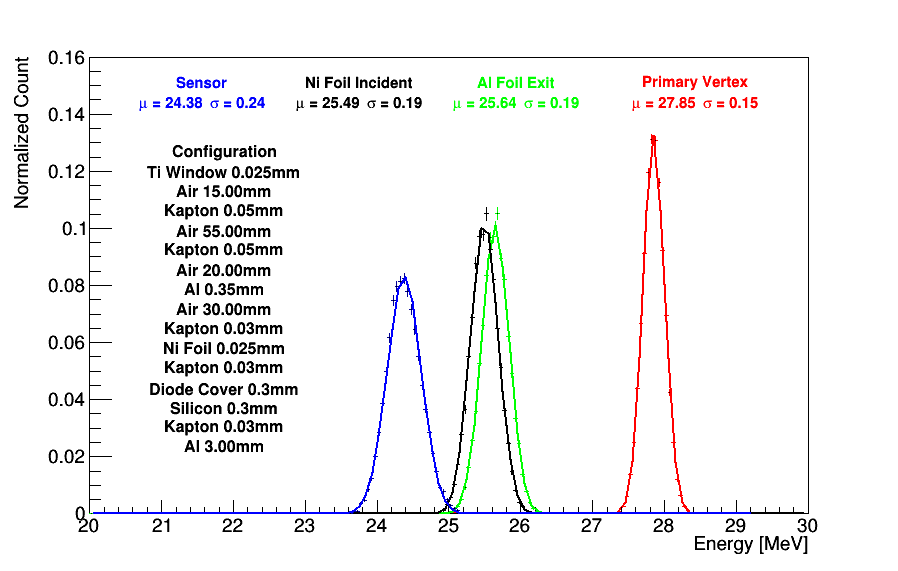}
  \caption{Geant4 simulation of the MC40 cyclotron beam-line showing the incident proton beam energy, the energy at the nickel foils, and the energy at the photodiodes.\label{fig:28MeV_incidentEnergy}}
\end{figure}

The irradiated sample consists of an aluminium plate with twelve slots
for diodes, mounted in pairs, as shown in
Fig.~\ref{fig:DiodeMount}. In front of each pair of diodes,
${}^{57}$Ni foils are installed and their activity after irradiation
is used to estimate the delivered proton fluence. Inside the
temperature controlled chamber, the sample and the foils, are placed
behind a $350\;\mu$m thick sheet of aluminium to block possible low
energy components of the beam. The energy of the proton beam when they
reach the sample, is estimated using a
Geant4-based~\cite{Agostinelli:2002hh, Allison:2016lfl} simulation, as
shown in Fig.~\ref{fig:28MeV_incidentEnergy}.

\subsection{IRRAD Proton Facility}

\begin{figure}[htbp]
        \centering
        \includegraphics[width = 0.8\linewidth]{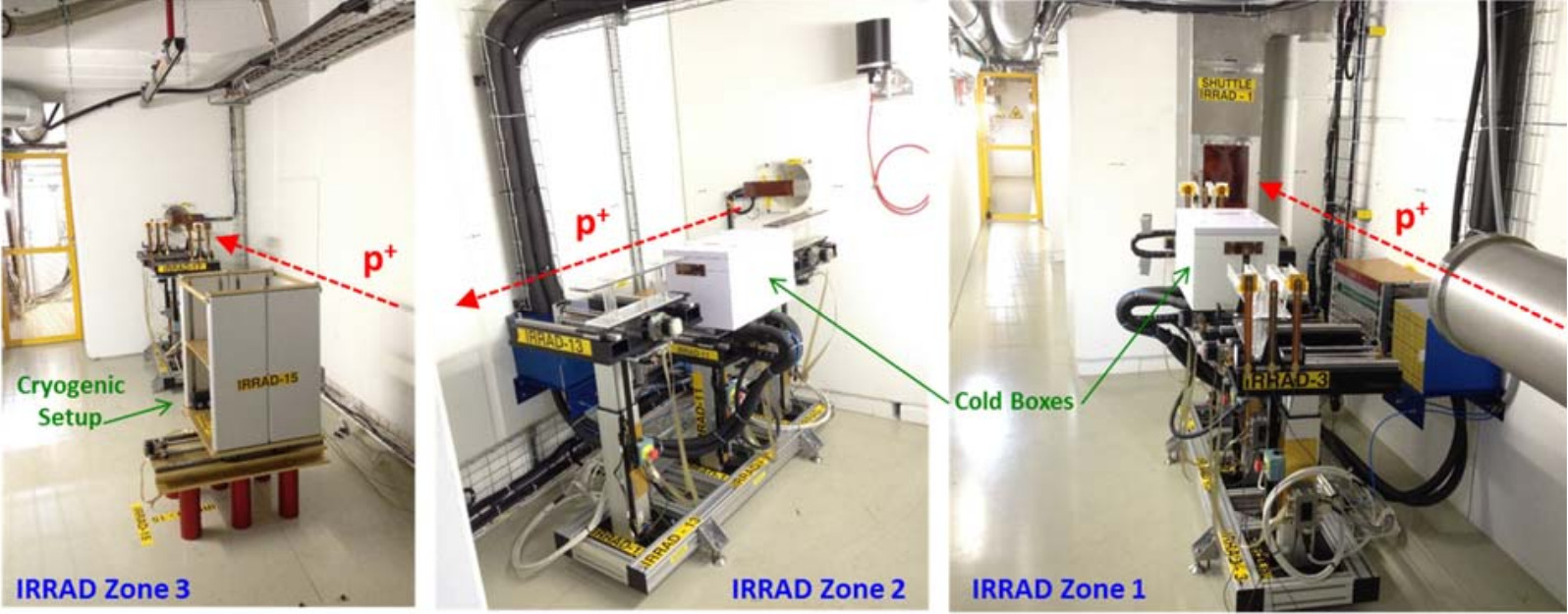}
        \caption{IRRAD proton facility experimental areas featuring three groups of remote-controlled tables installed along the
proton beam path, from Ref.~\cite{Gkotse:2015axt}.\label{fig:IRRAD_tables}}
    \end{figure}

The IRRAD proton facility at CERN utilises a primary proton beam with
an energy of $23\;$GeV, extracted from the proton
synchrotron~\cite{Cundy:2017ezr}. The facility employs the use of a
remote controlled stage to adjust the position of the sample, and an
isolated box for humidity and temperature control down to
approximately $-20\degree$C~\cite{Cindro:2014qca}. The proton fluence
determination for IRRAD is performed with aluminium foils.  Figure
\ref{fig:IRRAD_tables} shows an image of the IRRAD setup and the
remote controllable tables, which can move in the transverse and
azimuthal directions to align the sample with the beam. This setup
also allows for beam scanning, which can be done
automatically. Furthermore, there are three blocks of three tables
along the beam line, each separated by thick blocks of concrete.
    
For this study, both BPW34F photodiodes and FZ pad diodes were
irradiated to the same fluences for NIEL comparisons. Irradiations
take place at room temperature, and given the dose rate on the samples
no appreciable thermal annealing is expected to take place, in
particular when compared to the thermal annealing applied during the analysis procedure, see
\S~\ref{sec:Annealing}.

\subsection{Irradiation Center Karlsruhe}
\begin{figure}[htbp]
  \centering
  \includegraphics[width = 0.5\linewidth]{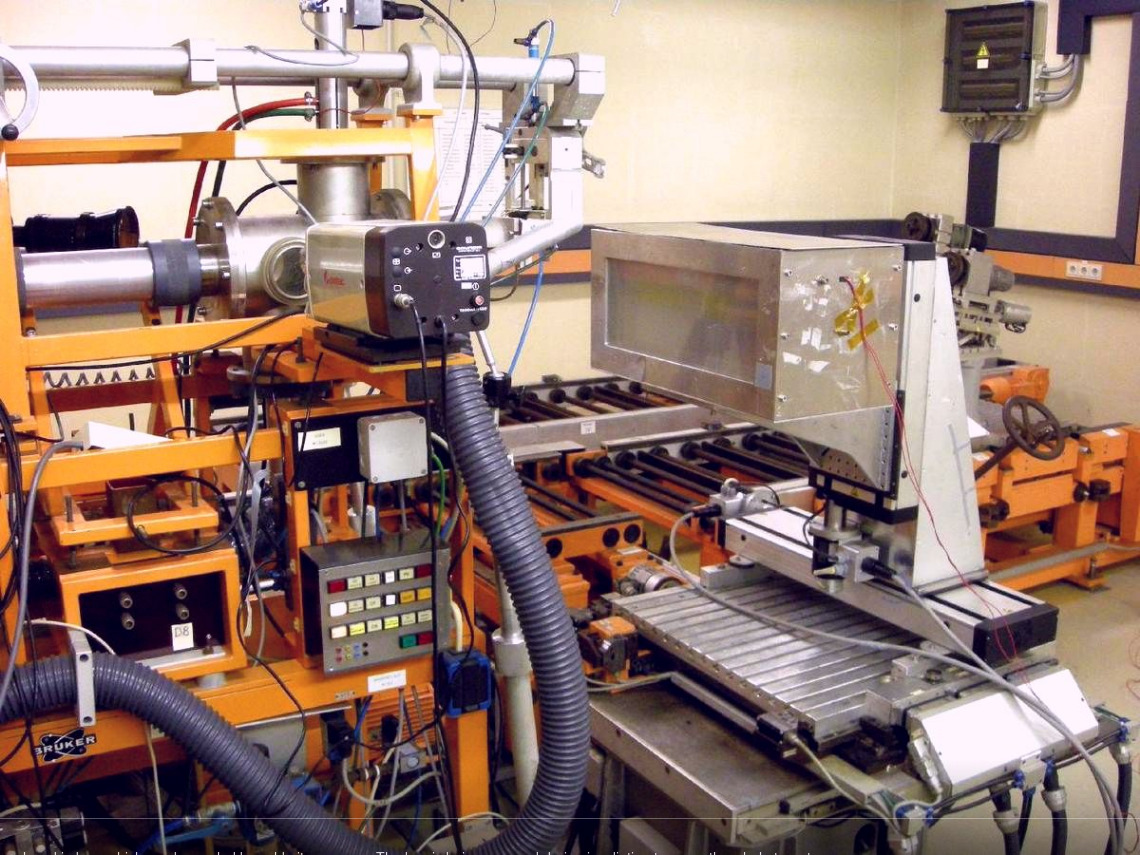}
  \caption{The Karlsruhe proton irradiation setup with the beam pipe visible on the left and the cold box on the right. During irradiation the box is moving to allow beam scanning over the whole area of interest.\label{fig:KIT_setup}}
\end{figure}

The Irradiation Center Karlsruhe~\cite{KITsite} accesses a compact
cyclotron operated by ZAG Zyklotron AG~\cite{ZAG}, a private-owned
company specialising in radioisotopes production for medicine and
engineering. The cyclotron accelerates protons to $25\;$MeV and for
this study the energy of the protons at the sample was measured to be
$23\;$MeV.  The set-up utilised for irradiations is shown in
Fig.~\ref{fig:KIT_setup}. Similarly to the MC40 cyclotron and the
IRRAD proton facility, the sample can be positioned within an
isolation box for humidity and temperature control down to
$-30\degree$C~\cite{KITsite}, and the fluences are calculated using
${}^{57}$Ni foils.

\section{Measurements}
\label{sec:measurements}
\subsection{Thermal Annealing Procedure}
\label{sec:Annealing}
Since the degree of thermal annealing significantly affects the
post-irradiation leakage current of photodiodes, all diodes where
thermally annealed for 80 minutes at $60\degree$C in accordance with
the guidelines of the RD50 collaboration. This ensured that
post-irradiation, all diodes possessed the same thermal history. The
process itself utilised a pre-heated oven, monitored with a NiCr-NiAl
thermocouple. For the MC40 cyclotron, due to the large number of
diodes tested, thermal annealing took place in two sets, with half of
the diodes in each set.

\subsection{Maximum Depletion Voltage}
\label{sec:MaxDep}
\begin{figure}[htbp]
  \centering 
  \subfigure[\label{fig:CVsetupRight}]{\includegraphics[width=.35\textwidth]{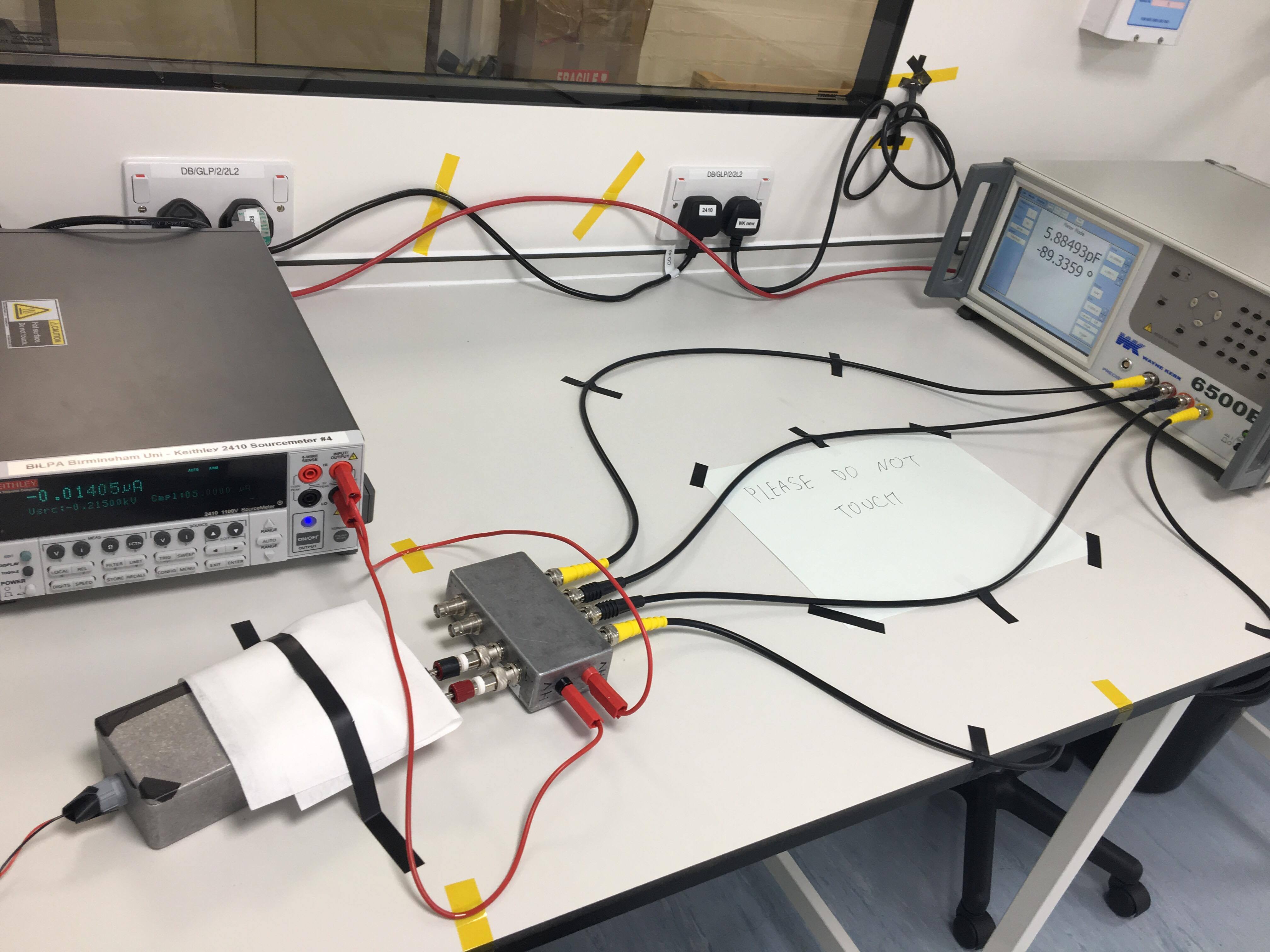}}
  \subfigure[\label{fig:CVsetupLeft}]{\includegraphics[width=.35\textwidth]{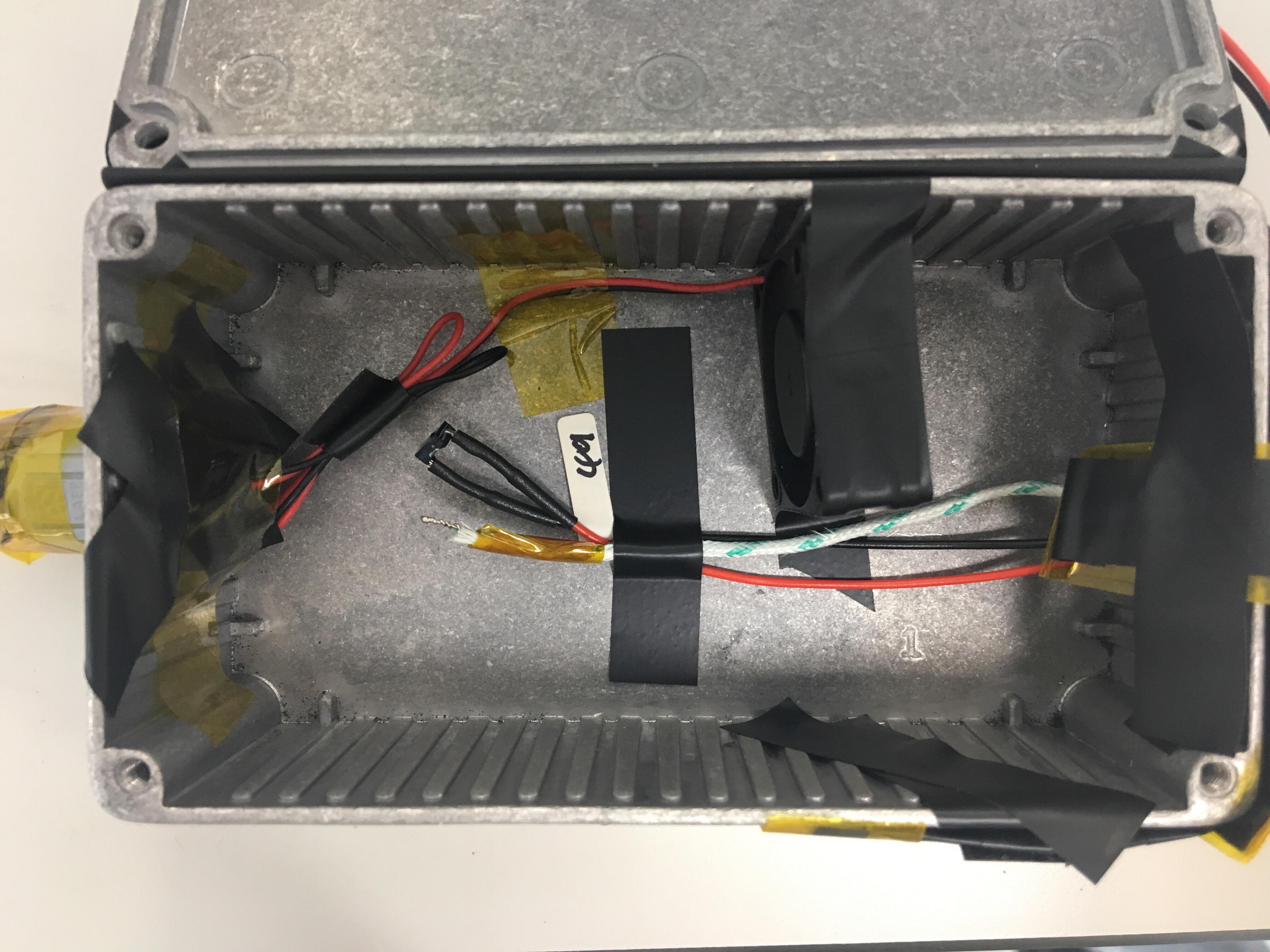}}
  \caption{\subref{fig:CVsetupRight} C--V measurement set-up, with the Wayne-Kerr 6500B
Precision Impedance Analyser, the Keithley 2410 Sourcemeter, and the aluminium box; \subref{fig:CVsetupLeft} Internal view of the aluminium box.\label{fig:CVsetup}}
\end{figure}

To estimate the value of the voltage where maximum depletion is
achieved, the diodes were placed inside an aluminium box for radiation
shielding, as shown in Fig.~\ref{fig:CVsetup}, alongside a fan for air
circulation. The system was then connected to a Wayne-Kerr 6500B
Precision Impedance Analyser via a junction box and four coaxial
cables for capacitance readings at 10~kHz; in accordance with RD50
guidelines. An external bias was supplied to the diodes by a Keithley
2410 Sourcemeter. The system was then trimmed to approximately zero
capacitance with the diode unconnected before each set of data was
taken.

\begin{figure}[htbp]
    \centering
    \includegraphics[width = 0.45\linewidth]{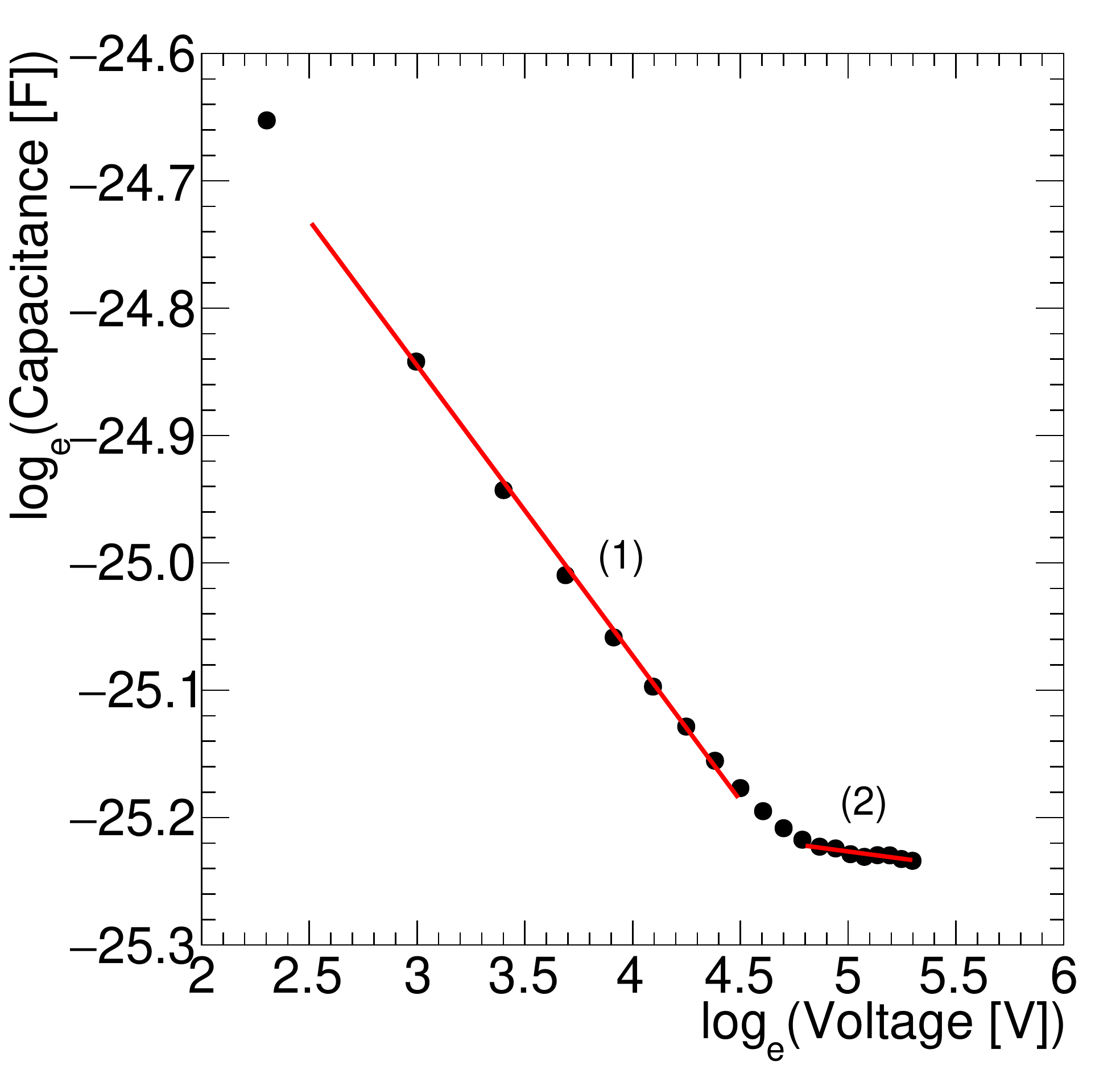}
    \caption{Capacitance as a function of reverse bias for a BPW34F photodiode irradiated to $4.33\times10^{11}$p/cm\textsuperscript{2}.\label{fig:Diode46_CV_PostAnneal_regions}}
\end{figure}

For a p-n junction, before full depletion, the capacitance is
inversely proportional to the square root of the voltage. Following
full depletion, capacitance becomes independent of the voltage. Using
this, it is possible to estimate the voltage at which a diode becomes
fully depleted, referred to as the maximum depletion voltage. Figure
\ref{fig:Diode46_CV_PostAnneal_regions} shows a plot of capacitance as
a function of voltage in logarithmic scales. In region (1), the diode
is not fully depleted, whilst in region (2), full depletion has been
achieved. In the latter region, the gradient is not zero as the
BPW34F diode does not contain a guard ring, and thus lateral depletion
is still occurring. By fitting the two regions linearly, the maximum
depletion voltage is estimated as the intercept of the two lines. This
estimate was performed for every diode following irradiation and
annealing.

\subsection{Quantifying Radiation Damage}
\begin{figure}[htbp]
        \centering
        \includegraphics[width = 0.5\linewidth, trim = {2cm 3.5cm 1.5cm 3.5cm}, clip = true]{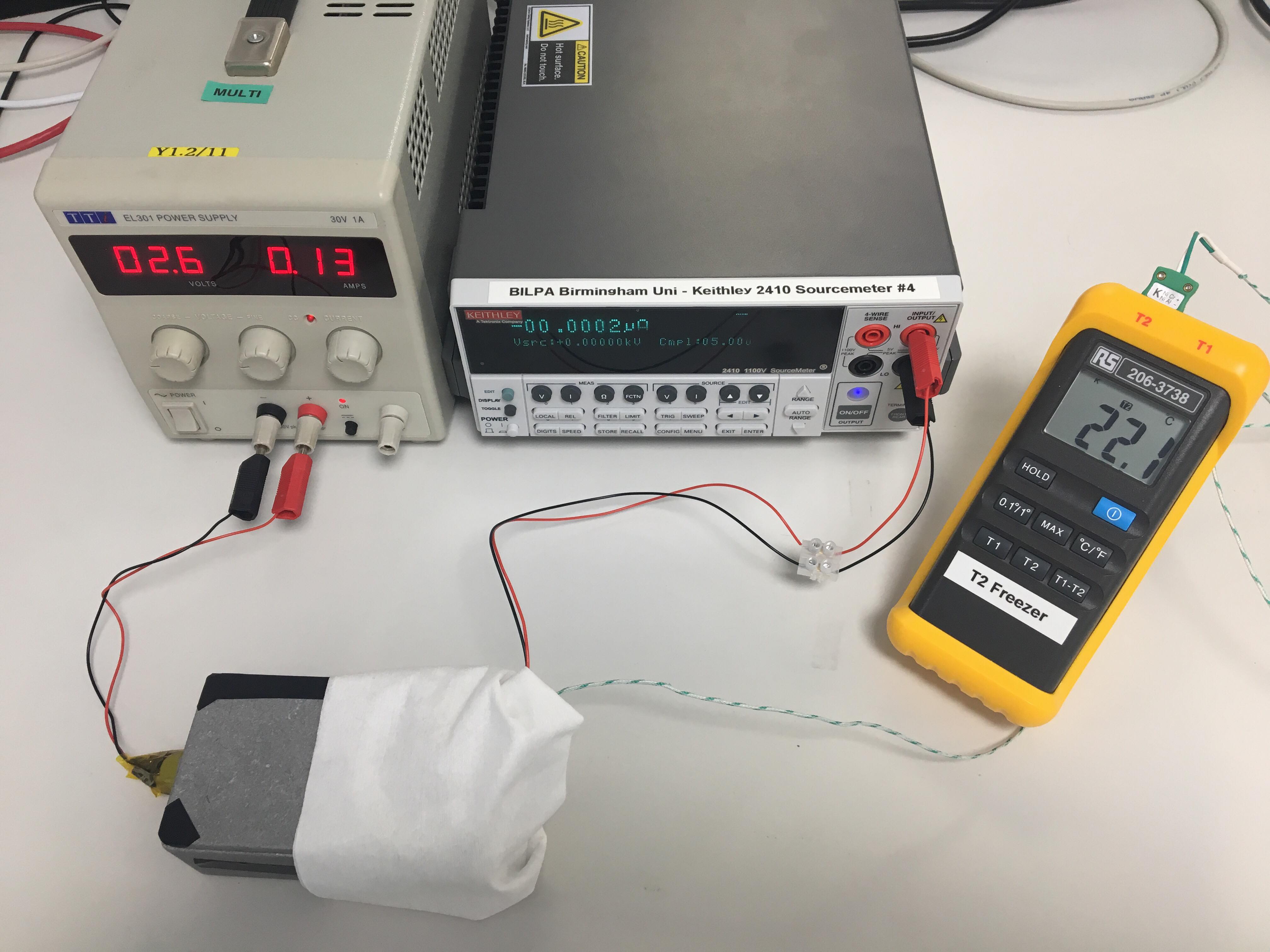}
        \caption{The setup utilised for I--V measurements.\label{fig:IVsetup}}
    \end{figure}

Figure \ref{fig:IVsetup} shows the experimental setup used for I--V
measurements. Similarly to the capacitance measurements, the diodes
were placed within an aluminium box alongside a fan. A Keithley 2410
source meter was used to apply a reverse bias across the diode, which
was then measured and displayed the corresponding current. A NiCr-NiAl
thermocouple was used to record the temperature within the box, being
placed close to the diode to obtain an accurate reading of its
temperature.

Although the ambient temperature, $T$, during data taking did not
deviate substantiall from $21\degree$C, to minimise any effects due to
the temperature dependence of the leakage current, all I--V curves
were normalised to the reference temperature, $T_R$, of $21\degree$C,
following RD50 recommendations. The formula used is given in
Eq.~\ref{eqn:tempscaling}, where $E_a$ is the activation energy of
silicon, which is closely related to the band gap energy of silicon,
and all other symbols have their usual meanings.

\begin{equation}\label{eqn:tempscaling}
  I(T_R) = I(T) \left(\frac{T_R}{T}\right)^2e^{-\frac{E_a}{2k_B}\left[\frac{1}{T_R}-\frac{1}{T}\right]}
\end{equation}
    
Over the temperature range relevant for this study, $E_a$ has a value
of $1.21\;$eV~\cite{Chilingarov_2013}. Post-irradiation, the leakage
current of the diodes increased proportionally to the incident fluence
due to induced defects in the silicon~\cite{Moll:1999kv}. Hence, the
change in leakage current pre- and post- irradiation can be used as a
measure of the degree of radiation damage. Assuming that the leakage
current scaled with the NIEL irrespective of the particle species and
energy, the change in leakage current can be used to determine the
hardness factor.

\subsection{Determination of Hardness Factors}
\label{sec:hfdet}
The change in leakage current pre- and post- irradiation, $\Delta I$,
as a function of fluence, $\phi$ is:

\begin{equation}
\label{eqn:Delta_I}
  \Delta I = \alpha l^2 w \phi
\end{equation}
    
where $\alpha$ is the current related damage rate, $l^2$ is the active
area of silicon, and $w$ is the width of the depletion region. For
BPW34F photodiodes, $l^2 = (0.265\times 0.265)$
cm\textsuperscript{2}~\cite{BPW34F,Ravotti:835408} and $w = 300\text{
}\mu$m~\cite{Ravotti:2008vcv}. From equation \ref{eqn:Delta_I}, and
the NIEL scaling hypothesis, it follows that the hardness factor can
be written as:
\begin{equation}\label{eqn:kappa}
\kappa = \frac{\phi _{neq}}{\phi},
\end{equation}
where $\phi _{neq}$ is the $1\;$MeV neutron equivalent
fluence. Combining equations \ref{eqn:Delta_I} and \ref{eqn:kappa},
the hardness factor can be obtained as:
\begin{equation}\label{eqn::kappaalpha}
  \kappa = \frac{\alpha}{\alpha _{neq}},
\end{equation}    
where $\alpha _{neq}$ is the current related damage rate for 1 MeV
neutrons. In this study, $\alpha _{neq} = (3.99\pm
0.03){\times}10^{-17}\;$Acm\textsuperscript{-1} was
used~\cite{Moll:1999kv}. Thus, by comparing the diode leakage current pre-
and post-irradiation as a function of fluence, the
hardness factor of the incident beam is calculated.

\begin{figure}[htbp]
  \centering 
  \subfigure[\label{fig:Diode47Left}]{\includegraphics[width=.45\textwidth]{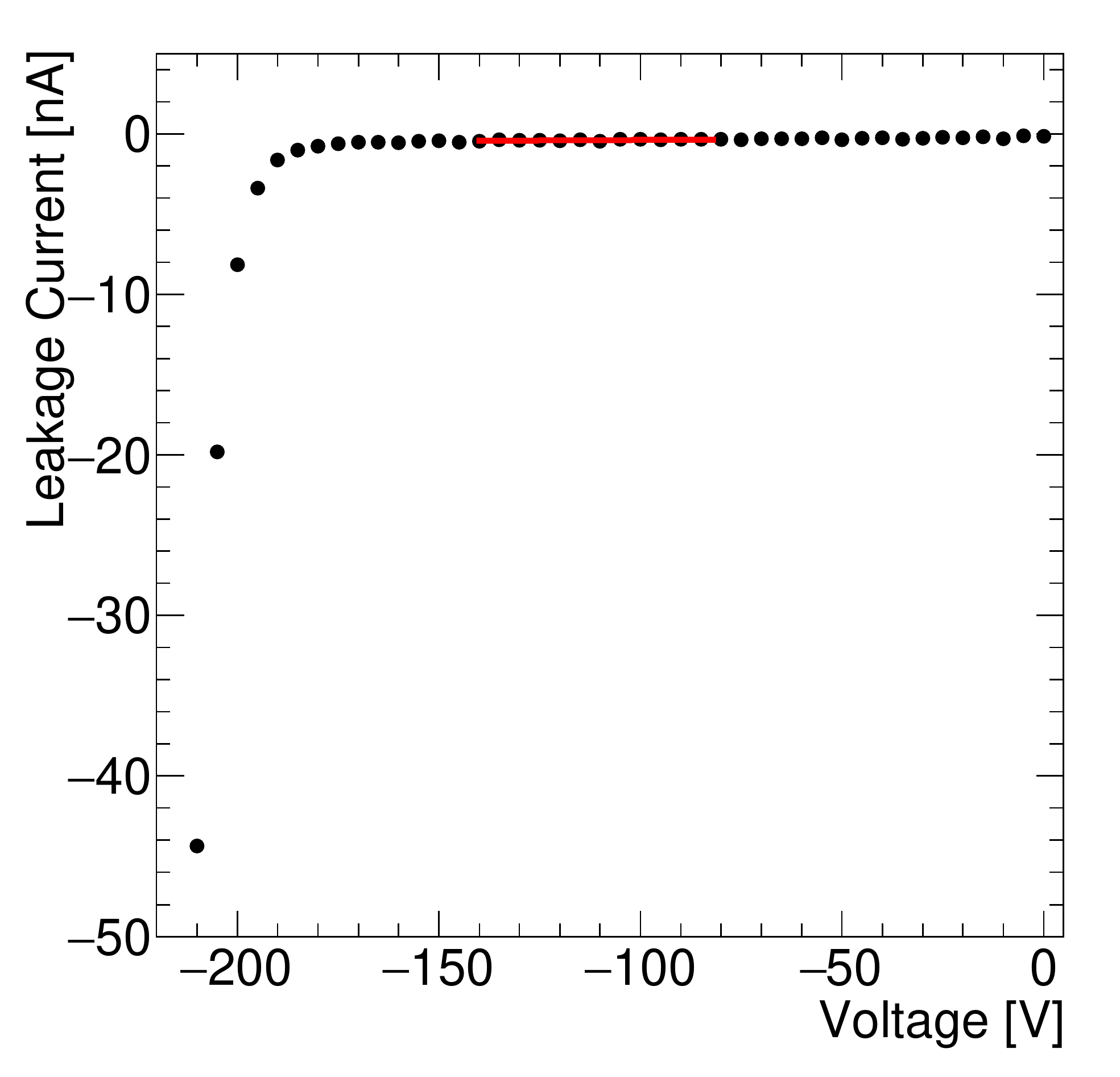}}
  \subfigure[\label{fig:Diode47Right}]{\includegraphics[width=.45\textwidth]{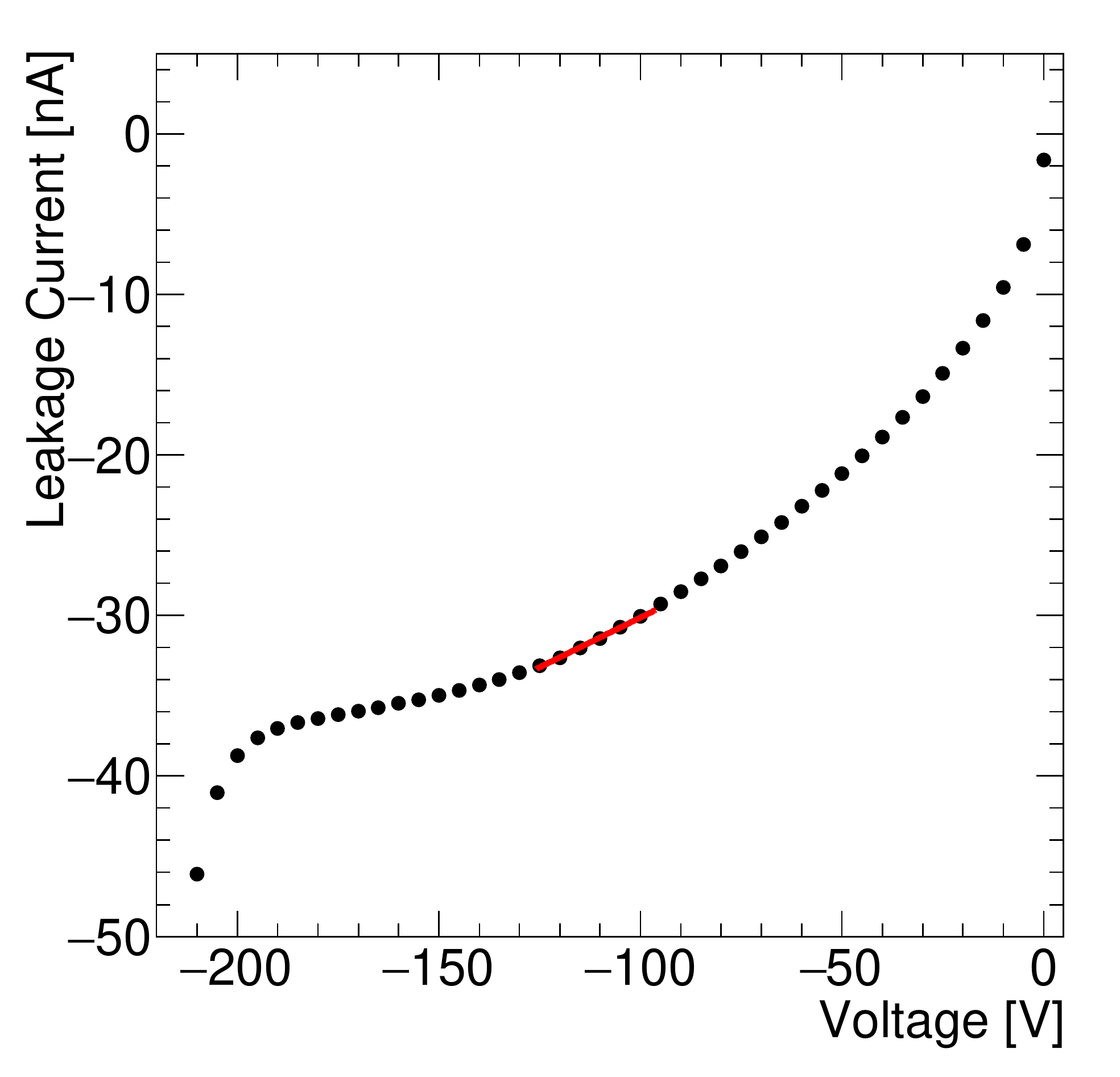}}
  \caption{I--V curves fit with a first order polynomial. \subref{fig:Diode47Left} Unirradiated diode; \subref{fig:Diode47Right} Following irradiation at $(1.64\pm 0.36)\times 10^{11}$pcm$^{-2}$ and thermal annealing.\label{fig:Diode47}}
\end{figure}
    
Figure~\ref{fig:Diode47} shows I--V curves for the same diode pre- and
post-irradiation and annealing. The data were fit with a first order
polynomial centred at the minimum voltage value for which the
depletion region is maximised, as determined from C--V
measurements. The change in leakage current, evaluated at this
voltage, was computed for each diode and plot as a function of
fluence.

\begin{figure}
  \centering
  \includegraphics[width=0.45\linewidth]{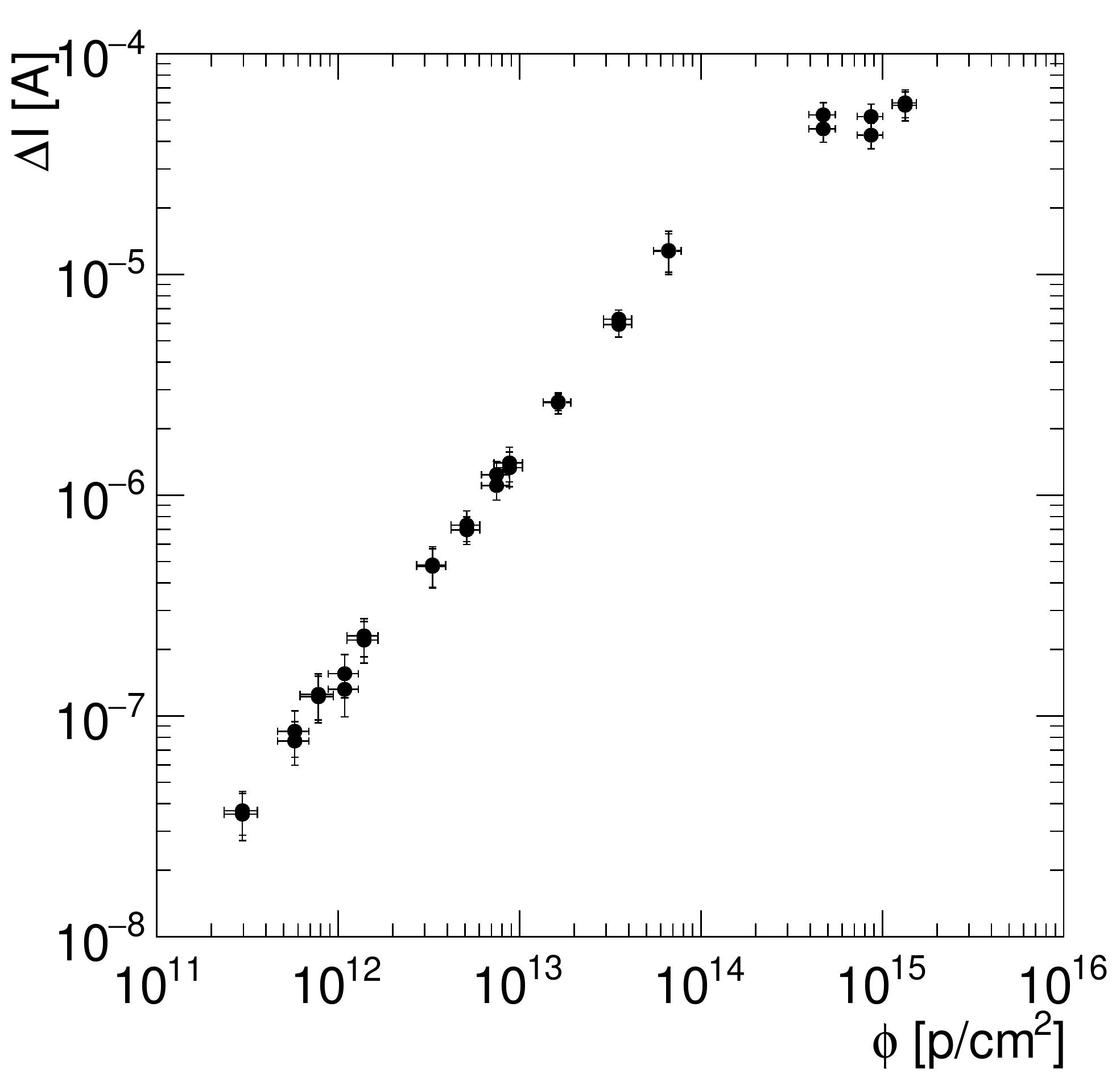}
  \caption{Change in leakage current as a function of fluence for the BPW34F diodes. The linearity of response up to approximately $10^{14}$p/cm\textsuperscript{2} is demonstrated.\label{fig:high_fluence}}
\end{figure}

Finally, the range of validity of Eq.~\ref{eqn:Delta_I} has been
considered in a dedicated study at the MC40 cyclotron. As shown in
Fig.~\ref{fig:high_fluence}, the change in leakage current is linear
as a function of fluence up to fluences of approximately
$10^{14}\;$pcm$^{-2}$. At low fluences, all charge carriers generated
by the radiation-induced defects are transported to the electrodes and
contribute fully to the leakage current (Shockley-Read-Hall
mechanism). At high fluences the high defect concentration results to 
charge carriers getting trapped. As a result, these are not 
transported throughout the sensor, which results to the observed plateau of the change in leakage current versus fluence.

\section{Results}
\label{sec:results}
\begin{figure}[htbp]
  \centering
  \subfigure[\label{fig:MC40_results}]{\includegraphics[width = 0.49\linewidth]{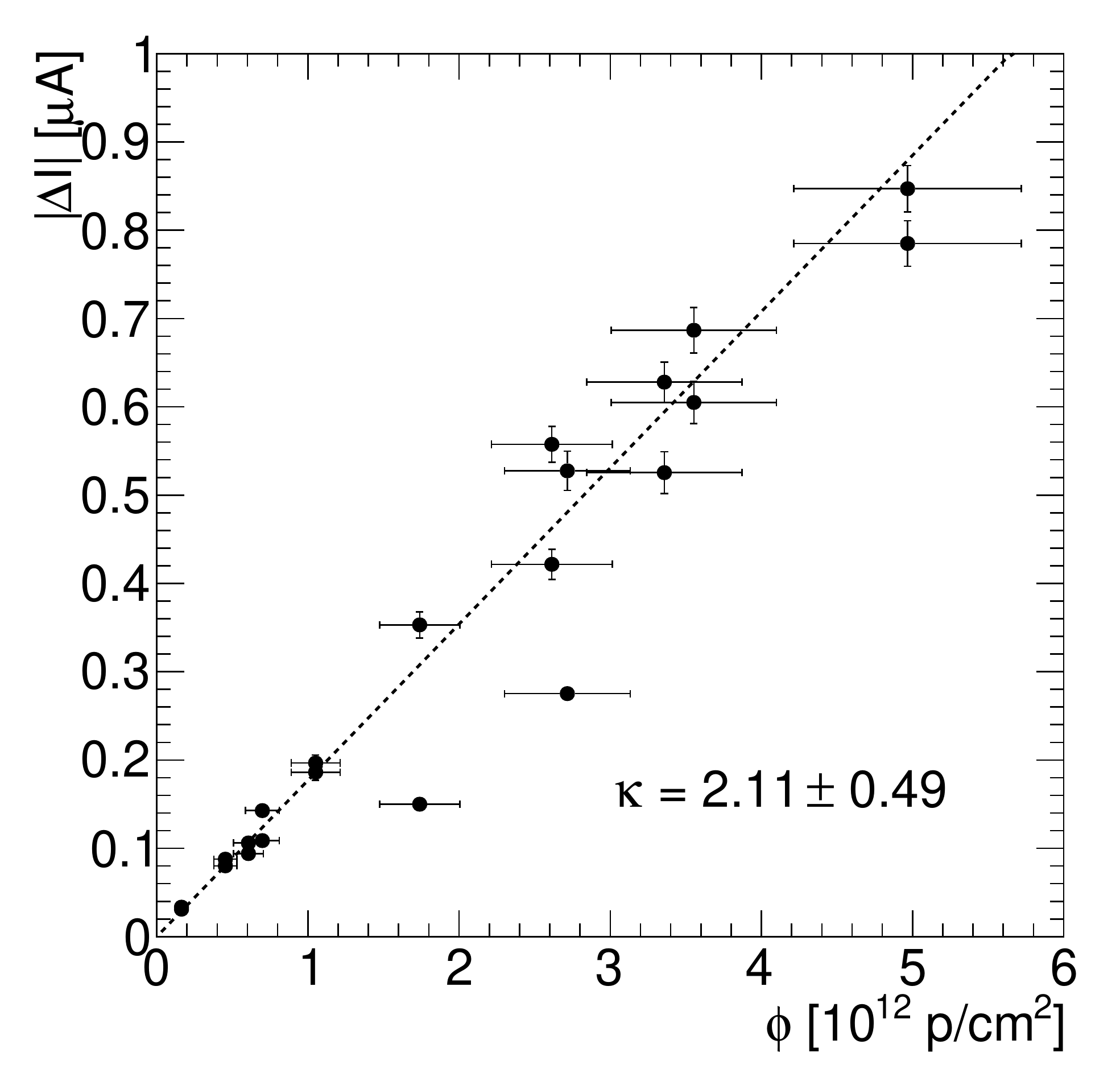}}
  \subfigure[\label{fig:KIT_results}]{\includegraphics[width = 0.49\linewidth]{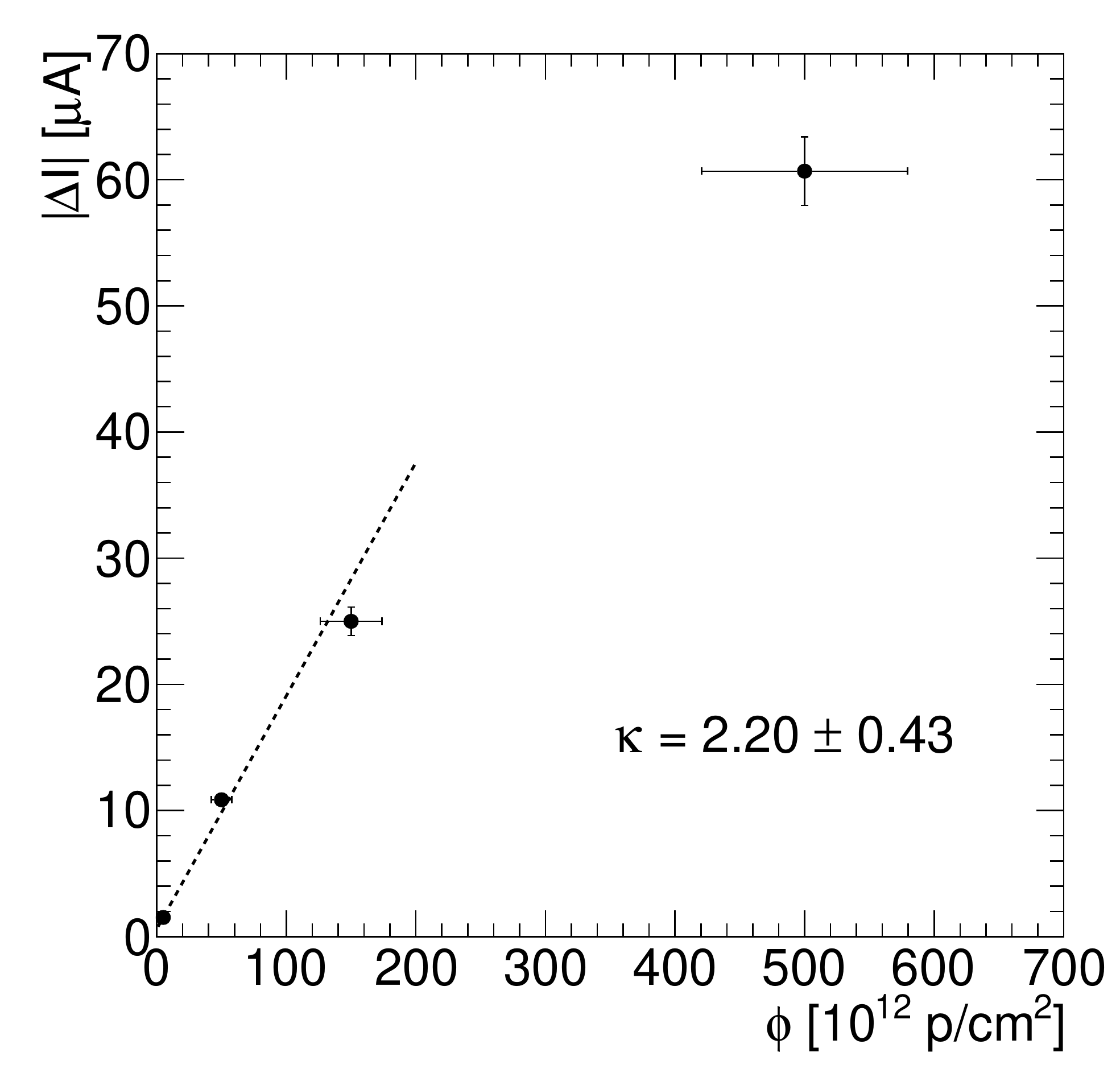}}
  \subfigure[\label{fig:IRRAD_results}]{\includegraphics[width = 0.49\linewidth]{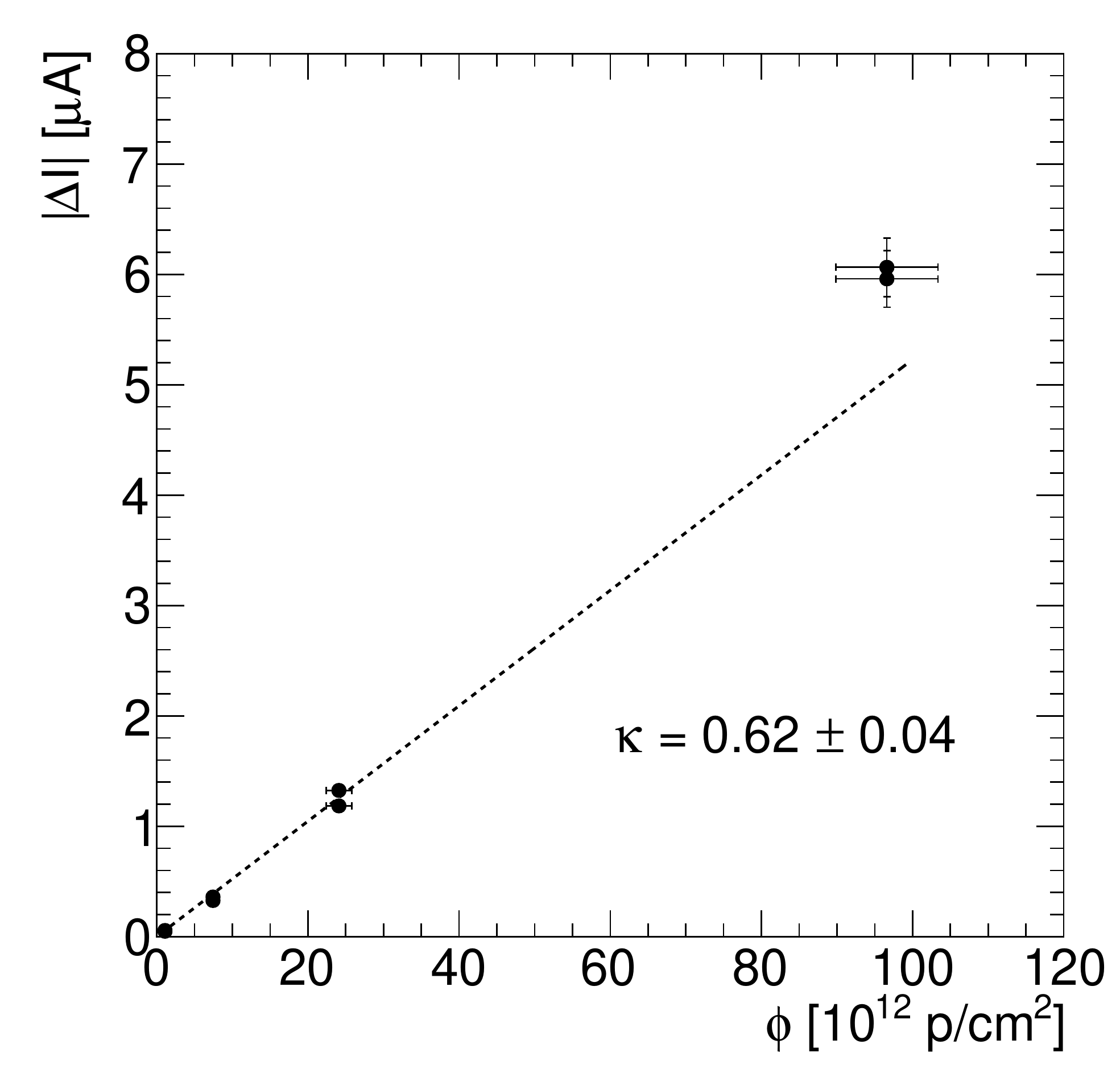}}
  \subfigure[\label{fig:FZ_results}]{\includegraphics[width = 0.49\linewidth]{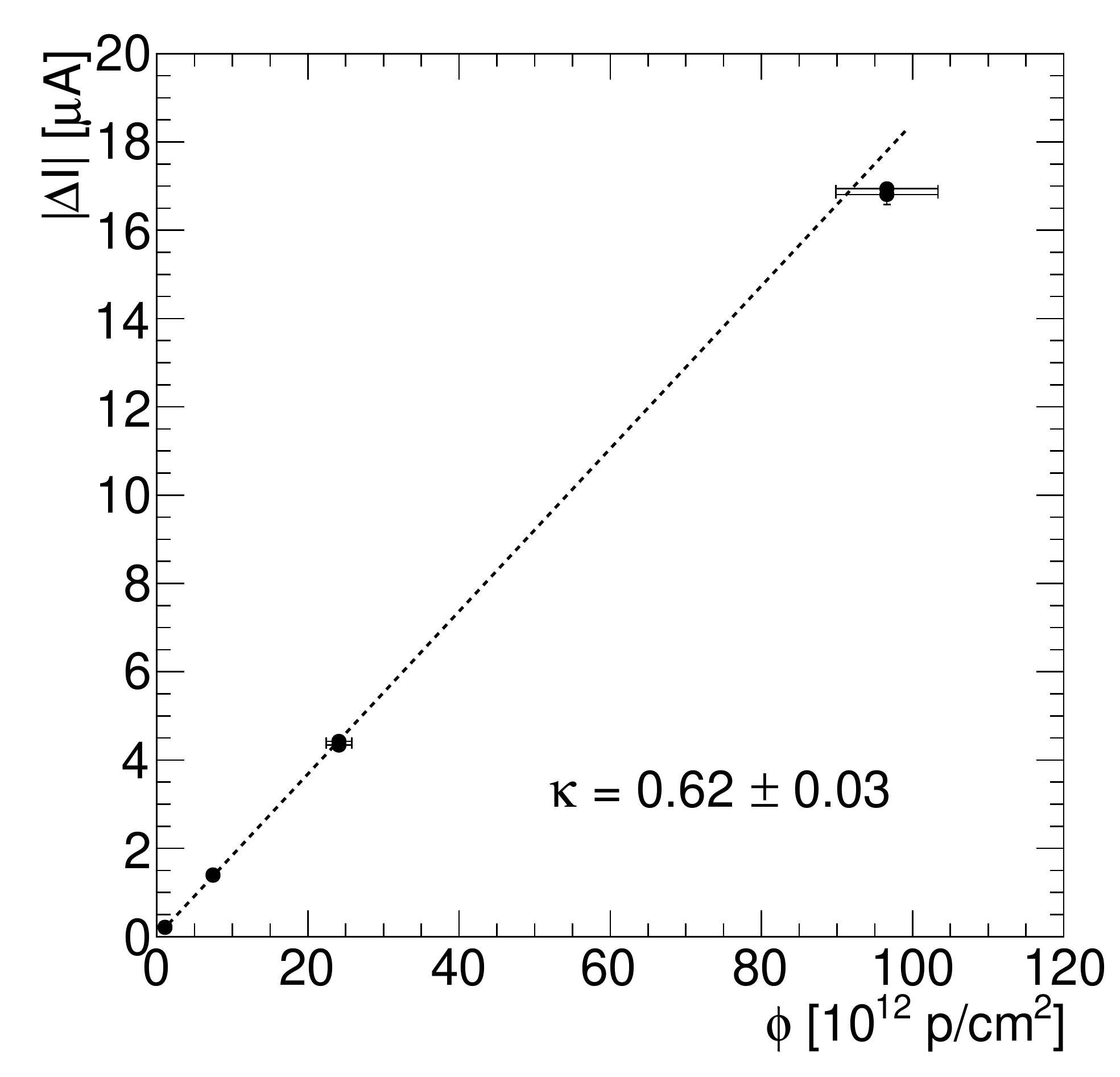}}
  \caption{Change in leakage current as a function of proton fluence
  for BPW34F photodiodes irradiated at \subref{fig:MC40_results} the
  MC40 cyclotron; \subref{fig:KIT_results} the Irradiation Center
  Karlsruhe; and \subref{fig:IRRAD_results} at the IRRAD proton
  facility; \subref{fig:FZ_results} FZ pad diodes irradiated at the
  IRRAD proton facility.}
\end{figure}

The measured change in leakage current as a function of the fluence
for BPW34F photodiodes, irradiated with $24\;$MeV protons at the MC40
cyclotron, are shown in Fig.~\ref{fig:MC40_results}. The data were fit
with a straight line, which was required to pass through the
origin. As there should be no change in leakage current, unless the
diode is irradiated. Nevertheless, leaving the intercept free in the
fit, did not significantly change the result. Using
equations \ref{eqn:Delta_I} and \ref{eqn::kappaalpha}, a hardness
factor value of $\kappa^{24\;{\rm MeV}} = 2.11\pm 0.49$ was
determined.
    
The measured $\Delta I$ as a function of fluence for a collection of
BPW34F and BPW34FS diodes irradiated with 23 MeV protons at the
Irradiation Center Karlsruhe is presented in
Fig.~\ref{fig:KIT_results}. For these data, the fit was not forced
through zero, as the pre-irradiation state for each diode was not
known. However, the change in leakage current between the pre- and
post-irradiated case is of several orders of magnitude, and so the
change in leakage current was approximated to the post-irradiated
leakage current following thermal annealing. 
Furthermore, given the discussion in Sec.~\ref{sec:hfdet}, the highest
fluence point of Fig.~\ref{fig:KIT_results} was omitted from the
fit. Applying equations \ref{eqn:Delta_I} and \ref{eqn::kappaalpha} as
before, a value of $\kappa^{23\;{\rm MeV}} = 2.20\pm 0.43$ was
determined.

Figure~\ref{fig:IRRAD_results} shows the change in leakage current as
a function of fluence for BPW34F photodiodes irradiated with $23\;$GeV
protons at the IRRAD proton facility. Again, the data were fit with a
first order polynomial that was required to go through the
origin. From this, a value of $\kappa^{23\;{\rm GeV}} = 0.62\pm 0.04$
was obtained.  In Fig.~\ref{fig:FZ_results} the corresponding
results for FZ pad diodes, irradiated to the same fluences at IRRAD,
are presented. A value of $\kappa'^{23\;{\rm GeV}} = 0.62\pm 0.03$ was
obtained from these data.

\begin{figure}[htbp]
  \centering
  \includegraphics[width = 0.45\linewidth]{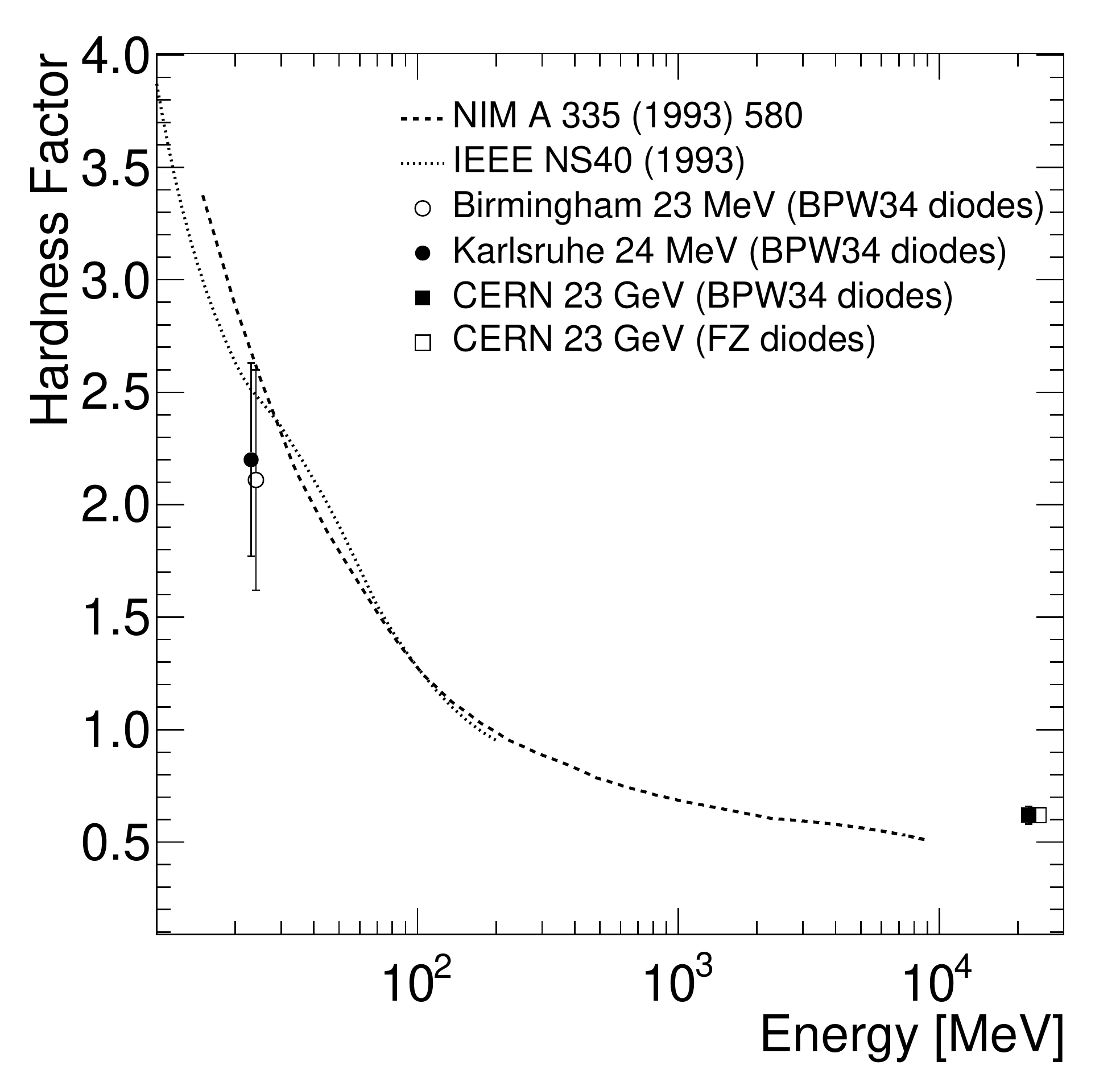}
  \caption{Measured proton hardness factors as a function of kinetic energy.\label{fig:measVsTab}}
\end{figure}

In Fig.~\ref{fig:measVsTab} the obtained measured values of the
hardness factor are compared with the tabulated values found in the
literature~\cite{vasilescu,Summers:1993,Huhtinen:1993np}. It is noted
that the significantly larger uncertainties on the hardness factor for
23 and 24 MeV protons stems from uncertainties in the dosimetry, and,
specifically, the uncertainty on the cross-section of the
$^{28}$Ni$(p,X)^{28}$Ni$^{57}$ process which is taken conservatively
to be approximately 20\% based on the spread of cross-section
measurements. Thanks to the higher proton energy, dosimetry at IRRAD
is performed using aluminium foils, with the respective cross-setion
uncertainty taken to be approximately
7\%~\cite{Curioni:2017gtq,Glaser:2006ay}. As a result, these
measurements could benefit from improved dosimetry, potentially
through the use of a, specifically designed, transmission ionisation
chamber that could operate at large beam currents.

In the future, it is of interest to supplement this work with
measurements at additional energies, for example the Cyclotron
RadioIsotope Center (CYRIC) at Tohoku
University~\cite{Nakamura:2015sva} and the Los Alamos Neutron Science
Center (LANSCE) facility~\cite{LANSCE} provide 70 and 800~MeV protons,
respectively. Another development could involve other particle species, for example the PSI
High Intensity Proton Accelerator (HIPA) provides irradiations with
300~MeV/c pions and the Jo\v{z}ef Stefan Institute TRIGA Mark II
reactor~\cite{AGHARA2006181,Snoj:2012dib} provides irradiations with
neutrons of a wide range of energies. An overview of the available
irradiation facilities worldwide is provided by a CERN-maintained
dedicated database~\cite{cerndb}.

\section{Conclusions}
\label{sec:summary}
By analysing the I--V and C--V characteristics of reverse biased
BPW34F photodiodes pre- and post-irradiation, the hardness factors,
$\kappa$, for three different proton energies were
determined. Utilising the University of Birmingham MC40 cyclotron, a
value of 
$2.1\pm 0.5$ 
for an energy of $24\;$MeV was obtained. By
undertaking a similar procedure, using the IRRAD proton facility, a
value of 
$0.62\pm 0.04$ 
for an energy of $23\;$GeV was measured. In
parallel, the corresponding measurements with FZ pad diodes irradiated
at IRRAD yielded a value of 
$0.62\pm0.04$. Using diodes irradiated at
the Irradiation Center Karlsruhe, for an energy of $23\;$MeV, a value
of 
$2.2\pm 0.4$ 
was found. All measured values agree with the hardness factors
currently in use at the facilities within one standard deviation. The
measurements for $23$ and $24\;$MeV protons have significantly larger
uncertainties with respect to the $23\;$GeV measurement. The origin of
this difference is in the dosimetry, which relies on nickel foils for
$23$ and $24\;$MeV protons, while aluminium foils are used for the
$23\;$GeV protons. In the future, improved dosimetry could be achieved
by using a transmission ionisation chamber, specifically designed for
high rate operation.

\section*{Acknowledgements}
The authors would like to acknowledge the help and support of the
operations teams at the Birmingham MC40 Cyclotron, the Irradiation
Center Karlsruhe, and CERN IRRAD, without which this work would not be
possible. This project has been supported from the European Research
Council (ERC) under the European Union’s Horizon 2020 research and
innovation programme (grant agreement No. 714893).  The
irradiation facilities have been supported as Transnational Access
Facilities by the European Union’s Horizon 2020 Research and
Innovation programme (grant agreement No. 654168).

\flushbottom

\bibliographystyle{ieeetr}
\bibliography{jinst-bham-paper}

\end{document}